\documentstyle[12pt,psfig]{article} 

\newcommand{\doe}
{This work was supported by the
Director, Office of Energy Research,
Office of High Energy
and Nuclear Physics,
Division of Nuclear Physics,
of the U.S. Department of Energy under Contract
DE-AC03-76SF00098.}
\newcommand{\adr}{Nuclear Science Division \& %
Center for Nuclear and Particle Astrophysics\\Lawrence Berkeley National %
Laboratory, Berkeley, California}
\newcommand{\auth}{Norman K. Glendenning}
\newcommand{\tit}{Spin-up of Solitary Pulsars: Signal of Phase Transition}

\newcommand{\eos}{equation of state~}

\newcommand{\eosp}{equation of state}
\newcommand{\Eos}{Equation of state}

\newcommand{\GR}{General Relativity~}

\newcommand{\GRp}{General Relativity}

\newcommand{\beqn}{\begin{eqnarray}}
\newcommand{\eeqn}{\end{eqnarray}}

\begin{document}

\begin{titlepage}
\begin{flushright}{LBNL-41309} \end{flushright}
\begin{center}
\begin{Large}
\tit {\footnote{\doe}}\\[3ex]
\end{Large}

\begin{large}
\auth\\[2ex]
\end{large}
\adr\\[2ex]
December 30, 1997 \\[2ex]
\end{center}

%%%%%%%%%% front cover picture ..........................
\begin{figure}[htb]
\vspace{-.6in}
\begin{center}
\leavevmode
\hspace{-.2in}
\psfig{figure=ps.t4,width=2.5in,height=3.0in}
\end{center}
\end{figure}

%\begin{center}
%%%{\bf PACS} 97.60.Gb,~97.60.Jd,~24.85+p \\[4ex]
%\end{center}

% FOR NOTATION OF WHERE PRESENTED USE THE FOLLOWING
\begin{quote}
\begin{center}
 {\bf Neutron Stars and Pulsars}\\
November 17 to 20, 1997\\
 Rikkyo Univiersity, Tokyo, Japan\\
 To be published in Universal Academy Press\\
 (Organizers: N. Shibazaki and  N. Kawai)
 \end{center}
 \end{quote}

\end{titlepage}

\begin{center}
\begin{Large}
\tit \\[5ex]
\end{Large}

\begin{large}
\auth\\[3ex]
\end{large}
\adr\\[3ex]
\end{center}

\begin{abstract}
A phase transition in the nature of matter in the core of a
neutron star, such as quark deconfinement or Bose condensation, can
cause the spontaneous spin-up of a solitary millisecond pulsar.
The spin-up epoch for our model lasts for $2\times 10^7$ years or 1/50
of the spin-down time (Glendenning, Pei and Weber 1997).
\end{abstract}

\section{Introduction}

Neutron stars have a high enough interior density as to make phase
transitions in the nature of nuclear matter a distinct possibility.
According to the QCD property of asymptotic freedom, the most
plausible is the quark deconfinement transition. According to lattice
QCD simulations, this phase transition is expected to occur in very
hot ($T\sim 200$ MeV) or cold but dense matter. 

Since neutron stars are born with  almost the highest density that
they will have in their lifetime, being very little deformed by
centrifugal forces, they will possess cores of the high density phase
essentially from birth if the critical density falls in the range of
neutron stars. However the global properties such as mass or size  of
a slowly rotating neutron star are little effected by whether or not
it has a more compressible phase in the core. In principle, cooling
rates should depend on interior composition, but cooling calculations
are beset by many uncertainties and competing assumptions about
composition can yield similar  cooling rates depending on other
assumptions about superconductivity and the cooling processes. 
Moreover,  for those stars
for which a rate has been measured, not a single mass is known. It is
unlikely that these measurements will yield conclusive evidence in the
present state of uncertainty.

Nevertheless, it may be possible to observe the phase transition in
millisecond pulsars by the easiest of measurements---the sign of
$\dot{\Omega}$. The sign should be negative corresponding to loss of
angular momentum by radiation. However, as we will show, a phase
transition, either first or second order, that occurs near or at the
limiting mass star, can cause spin-up during a substantial era
compared to the spin-down time of millisecond pulsars. We sketch the
conventional evolutionary history of millisecond pulsars with the
addition of the supposition that the critical density for quark
deconfinement falls in the density range spanned by neutron stars.

As already remarked, with this supposition, the star has a quark core
from birth but its properties are so little effected that this fact
cannot be discerned in members of  the canonical pulsar population. 
However, by some mechanism, usually assumed to be accretion from a
companion during the radio silent epoch following its $10^7$ year spin
down as a canonical pulsar, the neutron star may be spun up.
As it spins up, it
becomes increasingly centrifugally deformed and its interior density
falls. Consequently,  the radius at which the critical phase
transition  density occurs moves toward the center of the
star---quarks recombine to form hadrons. When accretion ceases, and if
the neutron star has been spun up to a state in which the combination
of reduced field strength and increased frequency turn the dipole
radiation on again, the pulsar recommences spin-down as a visible
millisecond pulsar.

During spin-down the central density increases with decreasing
centrifugal force. First at the center of the star, and then in an
expanding region, the highly compressible quark matter will replace
the less compressible nuclear matter. The quark core, weighed down by
the overlaying layers of nuclear matter is compressed to high density,
and the increased central concentration of mass acts on the overlaying
nuclear matter, compressing it further. The resulting decrease in the
moment of inertia  causes the star to spin up
to conserve angular momentum not carried off by radiation. The
phenomenon  is analogous to that of ``backbending'' predicted for
rotating nuclei  (Mottelson and Valatin 1960)
 and discovered in the 1970's (Johnson et.al. 1972, Stephens and Simon 1972)
(see Fig.\ \ref{nucleus}) In
nuclei, it was established that the change in phase is from a particle
spin-aligned state at high  nuclear angular momentum to a superfluid
state at low angular momentum.  The phenomenon   is also analogous to
an ice skater who commences a spin with arms outstretched. Initially
spin decreases because of friction and air resistance, but a period of
spin-up is achieved by pulling the arms in. Friction then
reestablishes spin-down. In all three examples, spin up is a
consequence of a decrease in moment of inertia beyond what would occur
in response to decreasing angular velocity.

\section{Calculation}
In our calculation, nuclear matter
was described in a relativistically covariant theory
(Garpman, Glendenning and Karant 1979,
Glendenning 1985, Glendenning and Moszkowski 1991) and quark matter
in the MIT bag model
(Chodos et.al. 1974). The phase transition occurs in a substance of
two conserved quantities, electric charge and baryon number, and must
be found in the way described in Ref. (Glendenning 1992). 
The moment of inertia
must incorporate all effects described above---changes in composition
of matter,  centrifugal stretching---and frame dragging,
all within the framework of \GRp. The expression derived by
Hartle is inadequate because it neglects these effects
(Hartle and Thorne 1968). Rather
we must use the expression derived by us  (Glendenning and Weber
1992a, 1992b).

\begin{figure}[tbh]
\vspace{-.5in}
\begin{center}
\leavevmode
\centerline{ \hbox{
\psfig{figure=ps.t1,width=2.5in,height=3in}
\hspace{.4in}
\psfig{figure=ps.t2,width=2.5in,height=3in}
}}
\begin{flushright}
\parbox[t]{2.4in} { \caption {Nuclear moment of inertia as a function of squared
frequency for $^{158}$Er, showing backbending in the nuclear case.
\label{nucleus}
}} \ \hspace{.4in} \
\parbox[t]{2.4in} {  \caption{\Eos~for the fist order
deconfinement phase transition described in the text.
\label{eos}
\label{eos_k300_y_h}
}}
\end{flushright}
\end{center}
\end{figure}

%%%%%%%%%%%%%%%%%%%%%%%%%%%%%%%%%%%%%%%%%%%%%

For   fixed baryon number we solve \GR for a star rotating at
a sequence of angular momenta corresponding to an \eos that describes
the deconfinement phase transition from  charge neutral nuclear matter
to quark matter. The \eos is shown in Fig. \ref{eos}
 The moment of inertia as a function of angular momentum
does not decrease monotonically as it would for a gravitating fluid of
constant composition. Rather, as described above, the epoch over which
an enlarging central region of the star enters the more compressible
phase is marked by spin-up (Fig.\ \ref{oi}).

%%%%%%%%%%%%%%%%%%%%%%%%%%%%%%%%%%%%%%%%%%%%%%%%%%%%%%%%%%

\section{Spin-up Era}
To estimate the duration of the spin-up, we solve the deceleration
equation for the star with moment of inertia having the behavior shown
in
Fig.\ \ref{oi}. 

 From the energy loss equation 
 \begin{eqnarray}
 \frac{dE}{dt} =
  \frac{d}{dt}\Bigl(\frac{1}{2} I \Omega^2 \Bigr) =
   - C \Omega^{4}
    \label{energyloss}
      \end{eqnarray}
        for  magnetic dipole radiation we find
\begin{eqnarray}
\dot{\Omega}= -\frac{C}{I(\Omega)}
  \biggl[1  + \frac{I^{\prime}(\Omega) \,
\Omega}{2I(\Omega)}
 \biggr]^{-1} \Omega^3\,.
   \label{braking2}
\end{eqnarray}
This expression reduces to the usual braking equation when the moment
of inertia is held fixed. The braking index is not constant but varies
with angular velocity and therefore time according to
  \begin{eqnarray}
  n(\Omega)\equiv\frac{\Omega \ddot{\Omega} }{\dot{\Omega}^2}
  = 3   - \frac{ 3  I^\prime \Omega +I^{\prime \prime} \Omega^2 }
    {2I + I^\prime \Omega}
     \label{index}
      \end{eqnarray}
      where $I^\prime \equiv dI/d\Omega$ and $I^{\prime\prime}
      \equiv dI^2/d\Omega^2$. This holds in general even for a star whose internal composition does not change with angular velocity (inconceivable). In particular one can see that for very high frequency, the derivatives will be largest and the braking index for any millisecond pulsar near the Kepler frequency will be less than the dipole value of $n=3$.

\begin{figure}[tbh]
\vspace{-.5in}
\begin{center}
\leavevmode
\centerline{ \hbox{
\psfig{figure=ps.t3,width=2.5in,height=3in}
\hspace{.4in}
\psfig{figure=ps.t4,width=2.5in,height=3in}
}}
\begin{flushright}
\parbox[t]{2.4in} { \caption { Moment of inertia of corresponding
to a change of phase.
Time flows from large to small
$I$.
\label{oi} 
}} \ \hspace{.4in} \
\parbox[t]{2.4in} { \caption {The time evolution of
the braking index plotted over one
decade that includes the epoch of the phase transition.
\label{nt}
}}
\end{flushright}
\end{center}
\end{figure}
The braking index is shown as a function of time in Fig.\ \ref{nt} 
An anomalous value endures for $10^8$ years corresponding
to the slow spin down of the pulsar and the corresponding slow
envelopment of a growing central region by the new phase. The two
points that go to infinity correspond to the infinite derivatives of
$I$ at which according to (\ref{braking2}), the deceleration 
vanishes.
They mark the spin-up era.
The actual spin-up lasts for $2\times 10^7$ years or 1/50 of the
spin-down time for this pulsar. This could be easily observed in a
solitary pulsar and would likely signal a phase transition.

\section{Summary}
We have found the spin-up phenomenon in a pulsar model for which the
change in phase occurs in a star near the maximum mass. The phase
transition can be first or second order as long as it is accompanied
by a sufficient softening of the \eosp. We have not quantified  the
meaning of ``sufficiently soft''.

If no pulsar is observed to produce the signal, little is learned.
Just as in nuclear collisions, failure to observe a signal does not
inform us that the deconfined phase does not or cannot exist.\\[1ex]

\doe\\[2ex]
References:\\[1ex]
\noindent Chodos, A., Jaffe,  R. L.,  Johnson, K., Thorne, C. B.
 and Weisskopf, V. F.
1974 Phys.\
  Rev.\ D  9 3471.

\noindent
Garpman, S. I. A.,  Glendenning, N. K. and Karant, Y. J.
1979,  Nuc.\ Phys.\  A322 382.

\noindent
Glendenning, N. K., 
1985,  Astrophys.\ J.\ 293 470.

\noindent
Glendenning, N. K. and  Moszkowski, S. A.,
1991, Phys.\ Rev.\ Lett.\  67 2414.

\noindent
Glendenning, N. K., 
1992, Phys. Rev. D,  46 1274.

\noindent
Glendenning N. K. and  Weber,  F., 
1992, Astrophys.\ J.\  400 647.

\noindent
Glendenning, N. K.  and  Weber,  F., 1994,
Phys. Rev. D  50 3836.

\noindent
Glendenning, N. K., Pei, S.  and  Weber, F.,
1997, Phys.\ Rev.\ Lett.\  79 1603.

\noindent
Hartle J. B.  and Thorne, S. K., 1968,
 Astrophys.\ J.\  153 807.

\noindent
Johnson, A.,  Ryde, H.  and  Hjorth, S. A.,
1972,  Nucl.\ Phys.\  A179 753.

\noindent
Mottelson, B. R.  and  Valatin,  J. G.,
1960, Phys.\ Rev.\ Lett.\  5  511.

\noindent
Stephens, F. S. and Simon, R. S., 1972,
 Nucl.\ Phys.\  A183  257.

\end{document}